# Multi-layer Thick Gas Electron Multiplier (M-THGEM): a new MPDG structure for high-gain operation at low-pressure

M. Cortesi,[1,a)] S. Rost,[1] W. Mittig,[1] Y. Ayyad Limonge,[1] D. Bazin,[1] J. Yurkon,[1] and A. Stolz[1]

[1]*National Superconducting Cyclotron Laboratory (NSCL), Michigan State University, East Lansing (MI) 48824, U.S.A*

The operating principle and performances of the Multi-layer Thick Gaseous Electron Multiplier (M-THGEM) is presented. The M-THGEM is a novel hole-type gaseous electron multiplier produced by multi-layer printed circuit board technology; it consists of a densely perforated assembly of multiple insulating substrate sheets (e.g., FR-4), sandwiched between thin metallic-electrode layers. The electron avalanche processes occur along the successive multiplication stages within the M-THGEM holes, under the action of strong dipole fields resulting from the application of suitable potential differences between the electrodes. The present work focuses on investigation of two different geometries: a two-layer M-THGEM (either as single or double-cascade detector) and a single three-layer M-THGEM element, tested in various low-pressure He-based gas mixtures. The intrinsically robust confinement of the avalanche volume within the M-THGEM holes provides an efficient suppression of the photon-induced secondary effects, resulting in a high-gain operation over a broad pressure range, even in pure noble gas. The operational principle, main properties (maximum achievable gain, long-term stability, energy resolution, etc.) under different irradiation conditions, as well as capabilities and potential applications are discussed.

## I. INTRODUCTION

Experimental high-energy physics (HEP) programs, such as the Large Hadron Collider (LHC), the Relativistic Heavy Ion Collider (RICH), the International Linear Collider (ILC) and others, are the main driving force for the design and rapid development of new and advanced detector concepts. Among the many innovations of crucial importance in recent decades, the introduction of Micro-Pattern Gaseous detectors (MPGDs) [1] is one example of the engineering undertakings of the particle physics community toward reliable, large sensitive area detectors with unprecedented spatial resolution, high-rate capability, operational stability and radiation hardness. The introduction of MPGDs opened a new era of state-of-the-art detector technologies, and they now represent the benchmarks for detector developments in future accelerator-based experiments.

The great progress and knowledge acquired in meeting the challenges of HEP experiments have benefited a wide range of scientific and industrial applications [2–4], including experimental nuclear physics/astrophysics [5–6], medical physics [7–9], rare event searches [10–12], homeland security [13–14], and promise many more important developments in the future. However, as the most modern MPGD concepts were predominantly conceived for satisfying the specific requirements of the HEP experiments, their implementation to other field of applications, characterized by different challenges and different operational conditions, is not always straightforward. An example is the low-pressure gas tracker for heavy ionizing particles, which requires detectors capable of delivering a large dynamic range and a constant gain over time. For some other important new classes of experiments, such as the active-target time-projection chamber (AT-TPC) with rare isotope beams [15-16], avoiding the addition of a quencher to the main gas component (typically $H_2$, $D_2$, $^3He$, $^4He$, etc.) is the preferable solution to maximize the reaction yield under study while minimizing the background. However, it is extremely difficult to attain stable high-gain operation of proportional gaseous detector under these demanding conditions (low pressure and no quencher), due mainly to the considerable photon-mediated secondary effects that lead to an early transition from proportional avalanche mode to the streamer mode (discharge).

This work focuses on the development of a novel MPGD structure, specifically designed to provide high-gain operation at low pressure, including pure noble gas, with the aim of heavy-charged particle tracking for next-generation rare-isotope experiments. The operational principle, main properties and possible applications of this new MPGD structure will be discussed.

## II. THE MULTI-LAYER THICK GASEOUS ELECTRON MULITPLIER

Inspired by the multi-cascade THGEM detector concept [17], in which each multiplier is an individual perforated Cu-coated foil separated from the other elements by a small gap (typically a few mm), the M-THGEM is made of a single, robust assembly comprising of several THGEM elements stacked together. The M-THGEM is produced by multi-layer printed-circuit-board (PCB) technique, consisting of mechanical drilling of alternate layers of copper and core material (i.e. FR-4, Kapton, Kevlar, etc.) laminated together. As performed on the conventional THGEMs, the production process terminates with chemical etching of a small rim (typically 100 μm) around the holes, on both outer copper-clad surfaces. Larger rims have the advantages of reducing the probability of discharges due to mechanical defects and higher gains [18], although rather large gain fluctuations under irradiation, due to the charging up of the insulating surface, have been reported in THGEM-based detectors [19]. Suitable PCB vias provide access for the voltage bias of the inner electrodes. The M-THGEM is a very robust multiplier element which can cover very large areas (above 1 $m^2$) without the need of unwieldly spacer frames, generally used with more fragile structures to avoid deformations caused by electrostatic forces [20].

In this work, we have investigated the operation and performance of M-THGEM with two- and three-insulator substrates (FR-4). Each substrate has a thickness of 0.6 mm, for a total M-THGEM thickness of 1.2 mm (two-layer) and 1.8 mm (three-layer). The diameter and pitch of the holes are 0.5 mm and 1 mm, respectively. Figure 1 shows a schematic drawing of a two-layer (part a) and a three-layer (part b) M-THGEM detector.

The operation principle of the M-THGEM is similar to other holes-type gas multipliers: upon the application of a suitable potential difference between the various electrodes, a strong electric dipole field is established across the holes. Electrons created in the gas volume (drift region), above the M-THGEM, converge into the holes and are multiplied by the gas avalanche processes; each hole acts as an independent proportional counter. Because of the extended volume of the holes, significantly larger than the electron mean free path, a high electron multiplication is achieved also at very low pressure. Examples of the field strength map for a two-layer and a three-layer M-THGEM, using a symmetrical potential difference (300 V) applied between each stage, are illustrated in figure 2a and figure 3a, respectively. The field maps were computed by means of finite-element ANSII Maxwell software [21], while electric field lines from the top of the drift volume were calculated using the Garfield package (ver. 9) [22].

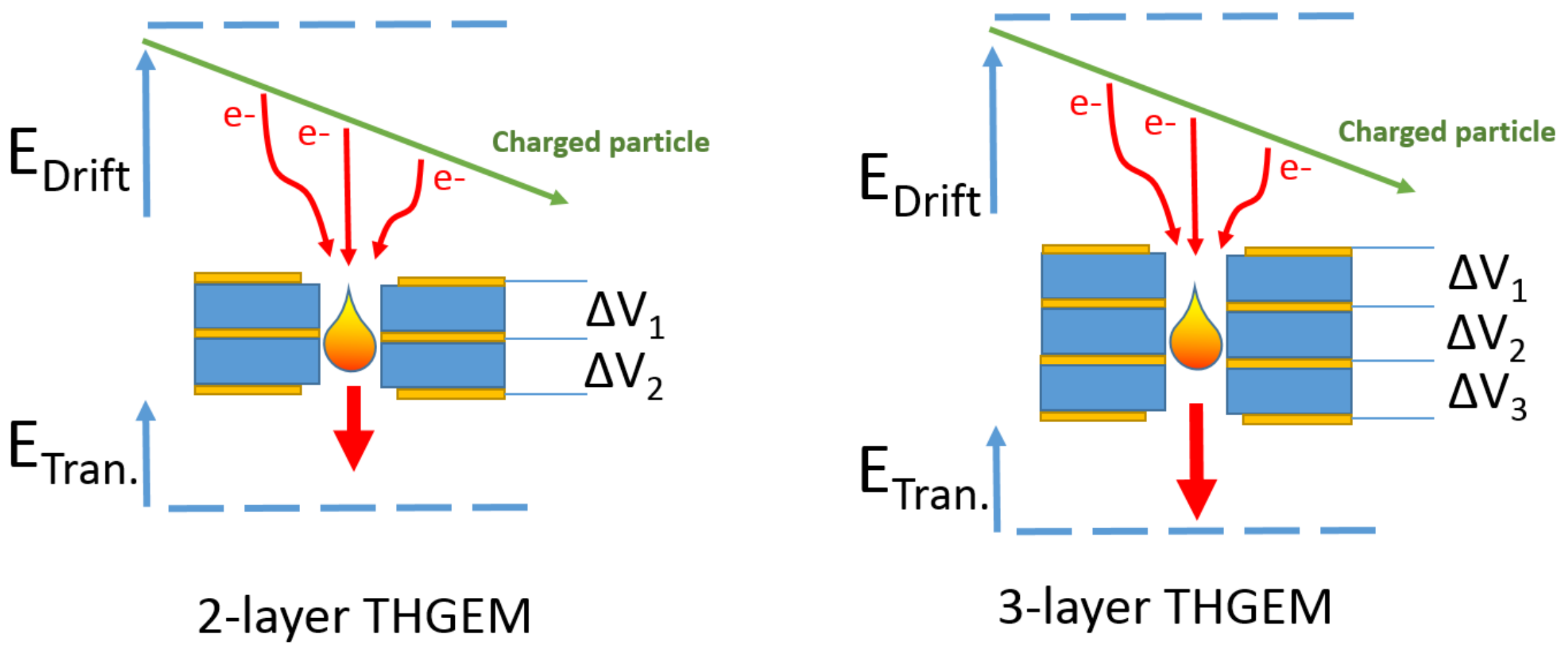

Figure 1. Schematic drawing of a two-layer (part a) and three-layer (part b) M-THGEM detector.

The M-THGEM can be operated either in a symmetric potential configuration, in which the same potential differences are applied between the various layers, as well as in an asymmetric configuration. In this latter arrangement, a slightly lower potential difference is applied across the first multiplier stage so that it acts as a positive ion ‘collector’, hence reducing the ion backflow to the drift gap. An example of the field map obtained in the case of an asymmetric bias applied to the two-layer M-THGEM is illustrated in figure 2b. The asymmetric configuration applied to the two-layer M-THGEM provides of better localized confinement of the avalanche within the innermost/bottom region of the M-THGEM holes. In this condition, the probability of feedback loops due to photon-mediated secondary effects, caused by the electroluminescence-induced emission of photoelectrons from the top electrode of the M-THGEM, is essentially suppressed; this is particularly important when the detector operates in pure noble gas, with no additional quenching component. As shown in figure 3, due to the strong electric field compacted along the central multiplication stage, the three-layer M-THGEM imparts an inherent true avalanche confinement within the holes also with a symmetric potential configuration, hence the three-layer M-THGEM is intrinsically ideal for pure noble gas applications.

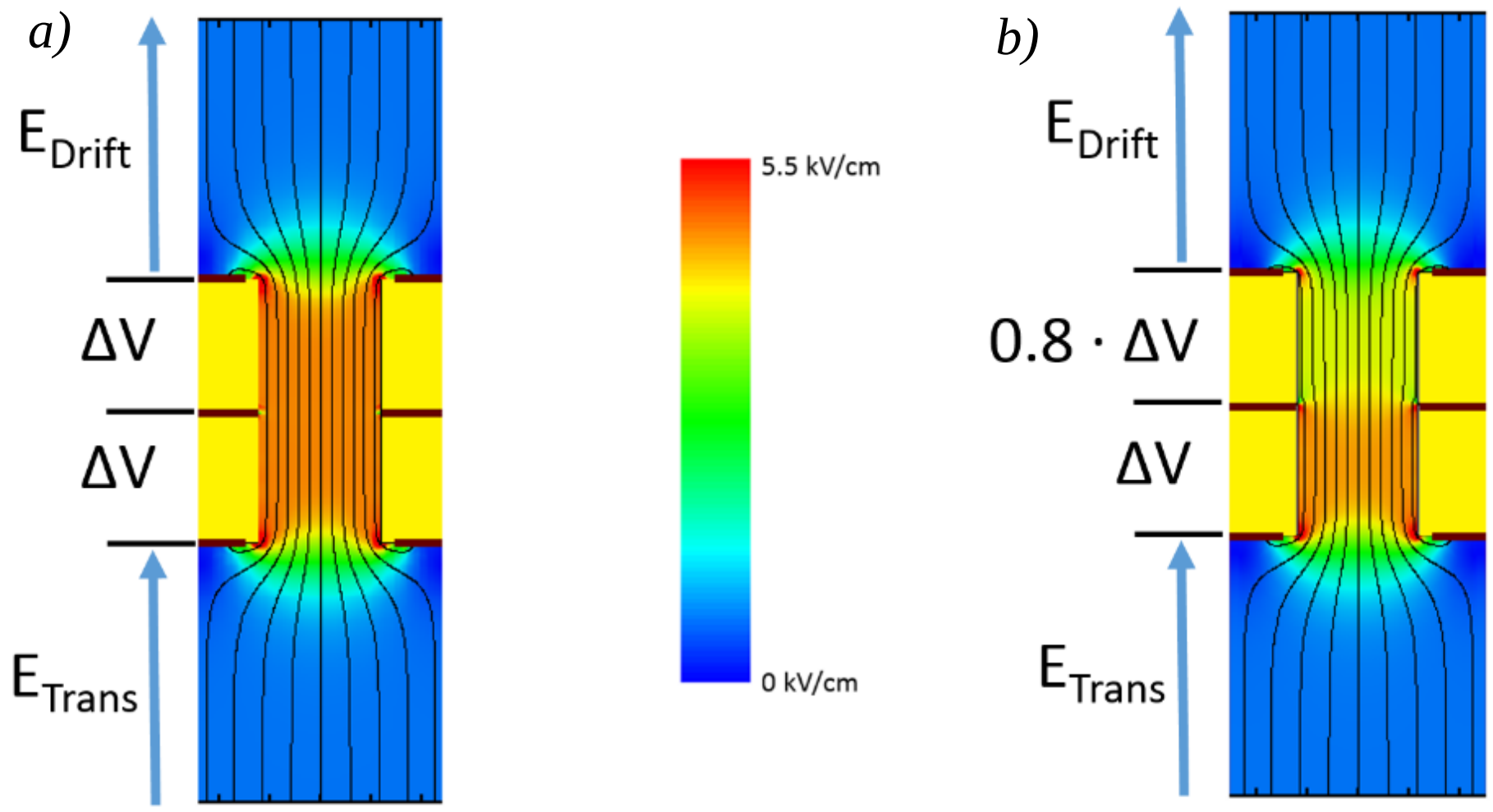

Figure 2. Electric field maps and field lines of a two-layer M-THGEM in the case of a symmetric (part a) and asymmetric bias (part b). In both cases $E_{Drift} = E_{Trans} = 0.5$ kV/cm and $\Delta V = 300$ Volt.

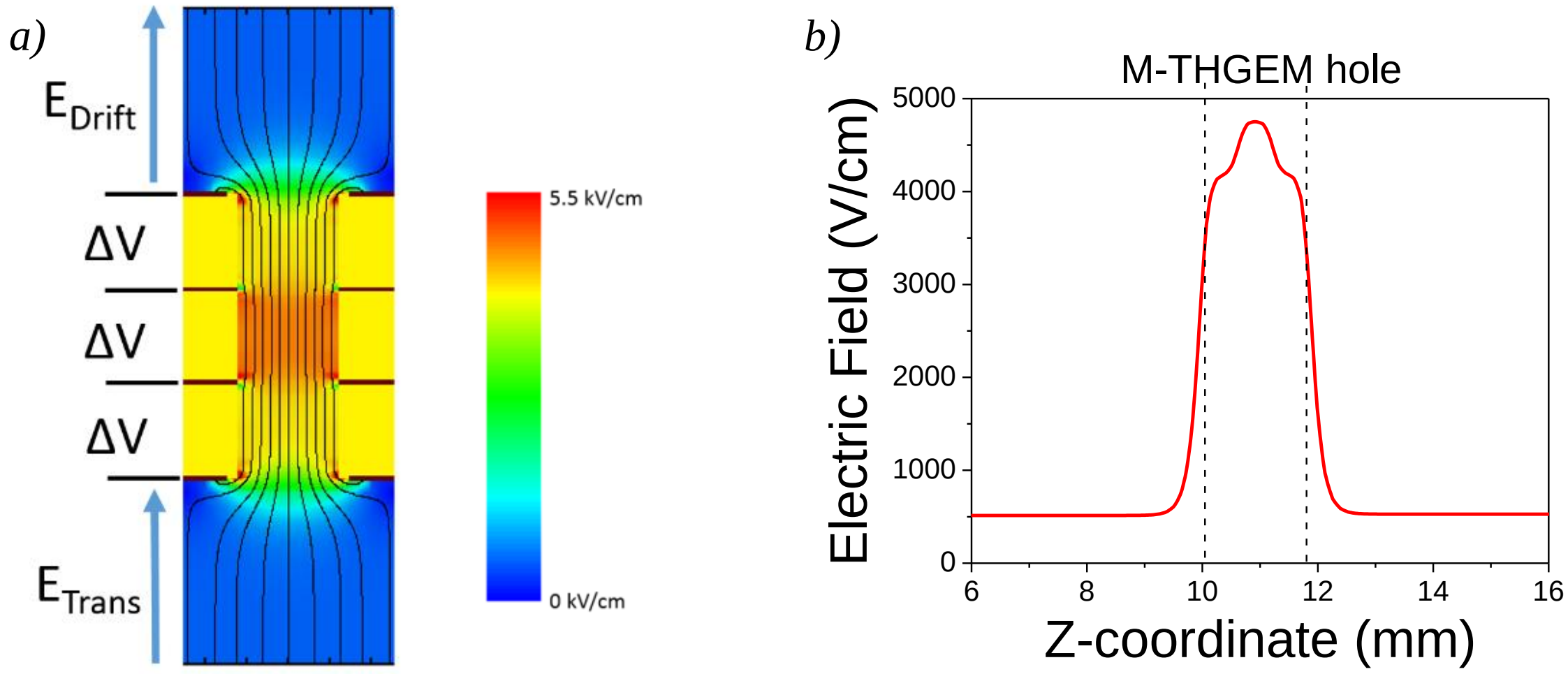

Figure 3. Electric field map (part a) and field lines along the hole axis (part b) of a three-layer M-THGEM, computed assuming $E_{Drift} = E_{Trans} = 0.5$ kV/cm and $\Delta V = 300$ Volt.

It is worth pointing out that in multi-cascade THGEM/GEM detectors, a sizable fraction of the avalanche charges created in a multiplier stage is lost due to electron collection either on the bottom electrodes of the multiplier itself, or on the top surface of the following one. On the contrary, the M-THGEM has the unique capability to transfer efficiently all the charges produced in one stage to the successive one without loss. As a result, the M-THGEM detector provides a lower-voltage, high-gain operation less prone to sporadic discharges and a lower probability of spark-induced damages.

Finally, an additional potential advantage provided by the M-THGEM is a possible reduction of the effective avalanche gain variation over time, due to the charging up of the insulator surface. Charging up effects are present in all gaseous detectors enclosing dielectric materials, caused by charges (electrons and ions) collected on the dielectric surfaces during the avalanche process. These charges modify the field geometry within the multiplier structure over time (several hours in THGEMs), with severe consequences on some essential functional parameters, such as the electron transparency and the effective gain. In the case of the M-THGEM detector, the inner electrodes stabilize the geometry and strength of the electric dipole field, resulting into an effective attenuation of charging up effects. This is particularly relevant for long hole-type structures, necessary for low-pressure operation, since the large dielectric surface exposed to the avalanche may capture and collect a substantial amount of charges, causing significant long-term variation of the effective gain.

## III. PERFORMANCE

### A. Two-layer M-THGEM

A preliminary basic study of the response of the two-layer M-THGEM detector (single and two-cascade configuration) was performed on small (10x10 cm$^2$) active area prototypes, in He-based mixtures and for a wide range of pressures (from 150 torr to 760 torr). The top surface of the first cascade M-THGEM was illuminated with high-intensity UV-light (see schematic drawings in figure 4a and figure 5a). The current collected at the readout was measured with a high-resolution pico-ammeter.

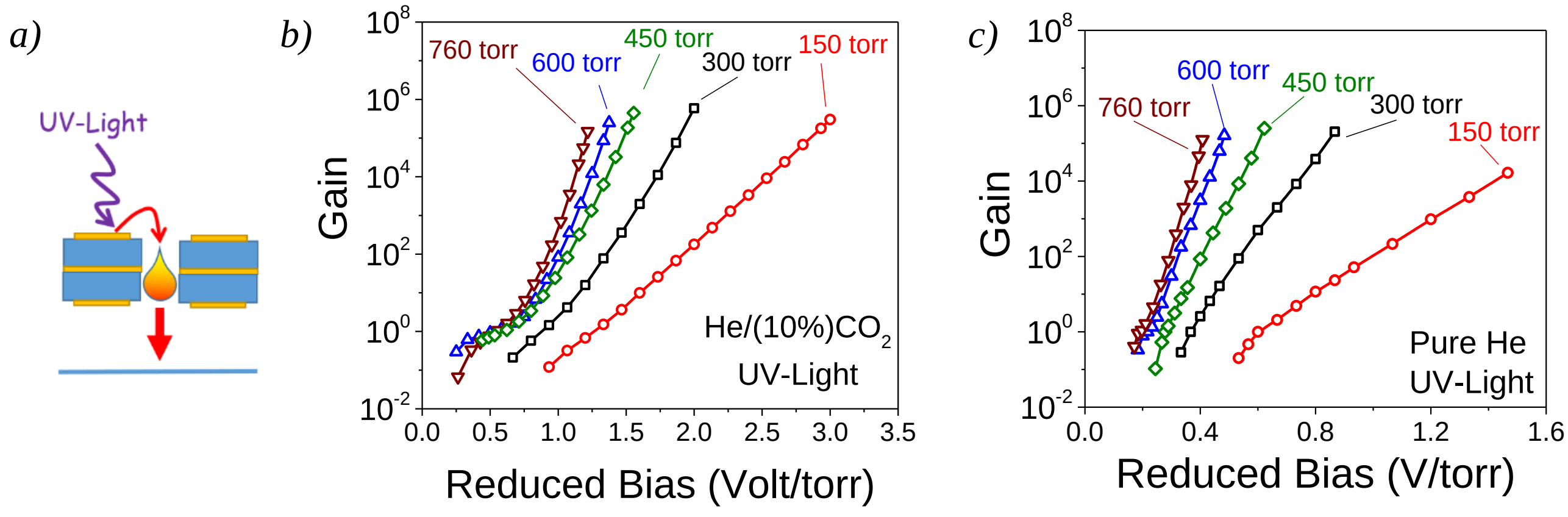

Figure 4. Single-photoelectron gain for a single two-layer M-THGEM (schematically drawn in part a), as function of the reduced bias (Volt/torr), in He/(10%)$CO_2$ (part b) and in pure He (part c).

Figure 4 depicts effective gains plotted as a function of the reduced bias (V/torr), symmetrically applied across each multiplication stage of the two-layer M-THGEM, respectively for He/(10%)$CO_2$ (part b) and in pure He (part c). Except for a minor drop observed at very low pressure in pure He, the detector reaches a more or less constant maximum gain of the order of $10^5$ over the full range of investigated pressures, in both gas mixtures. Figure 5 illustrates similar gain measurements performed on a two-cascade two-layer M-THGEM, under the same experimental conditions as for the single two-layer M-THGEM data, shown in figure 4. The two-cascade detector achieves a higher maximum gain of $10^6$, even at very low pressure (down to 150 torr), in both He/(10%)$CO_2$ (part b) and pure He (part c). Notably, there is no difference in terms of maximum achievable gain and spark probability between operation in pure and quenched He gas, as a result of the efficient suppression of the photon-mediated feedback effects. Figure 6 depicts a comparison between gain curves for single- and double-cascade two-layer M-THGEM detectors, measured in pure He at 150 torr (red graph) and 450 torr (green graph).

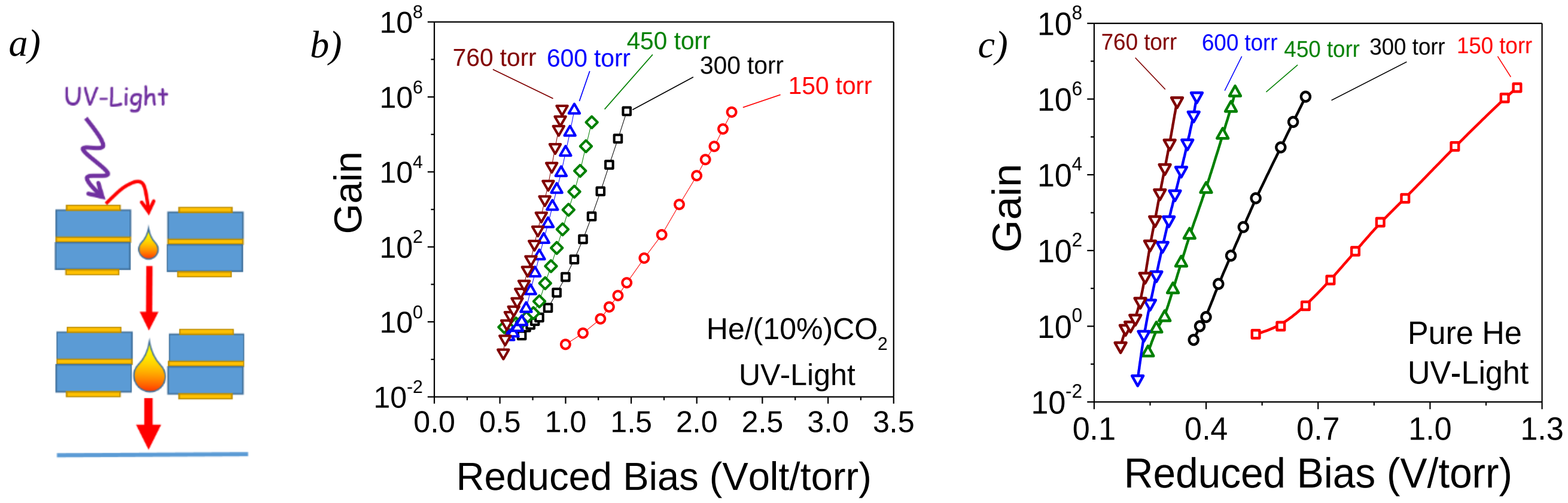

Figure 5. Single-photoelectron gain for a two-cascade two-layer M-THGEM (schematically drawn in part a), as function of the reduced bias (Volt/torr), in He/(10%)$CO_2$ (part b) and in pure He (part c).

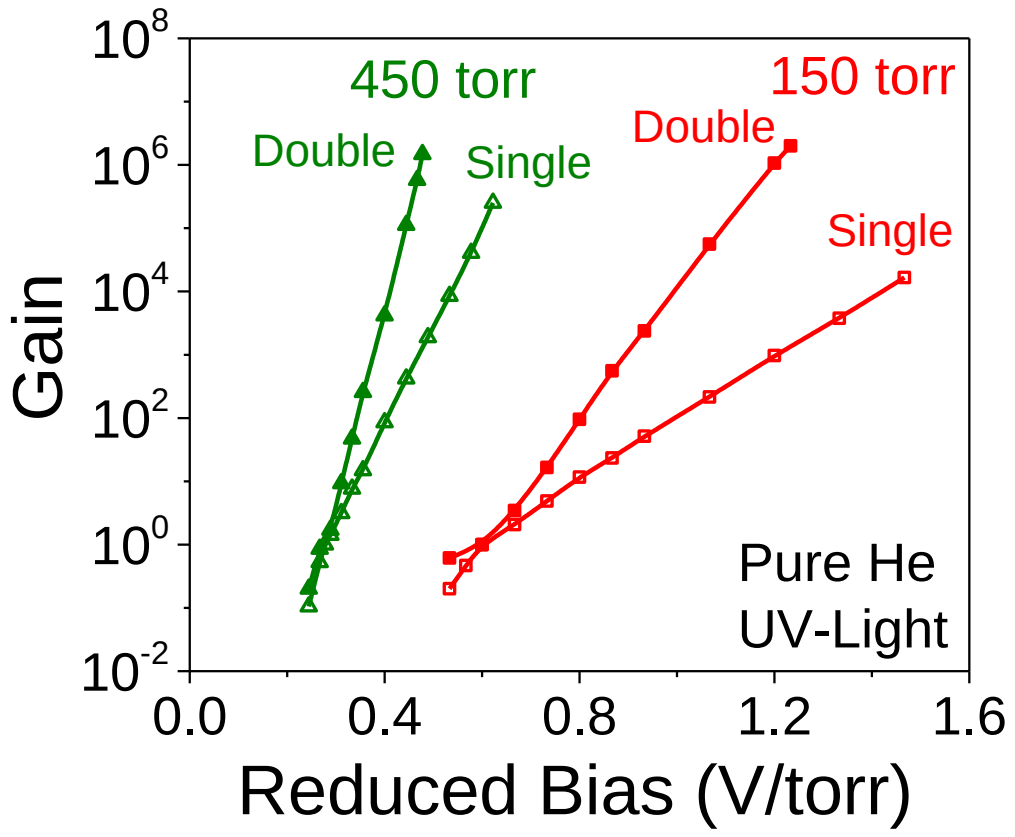

Figure 6. Comparison of single-photoelectron gain curves for single- and double-cascade two-layer M-THGEM detectors, measured in pure He at 150 torr (red graph) and 450 torr (green graph).

### B. Three-layer M-THGEM

The performance investigation of the three-layer M-THGEM (single element) detector was carried out on a large-area circular electrode, with a diameter of 25 cm and a total thickness of 1.8 mm. The electrode was mounted on top of the readout plate of the Prototype Active-Target Time Projection Chamber (pAT-TPC) [23], developed at the NSCL. The pAT-TPC consists of a cylindrical field cage measuring 50 cm in length and 28 cm in diameter, closed at one end by the readout plane, and on the other end by a cathode plate equipped with a thin entrance window (3.6 μm thick para-aramid film). Figure 7 shows the schematic drawings and a photograph of the M-THGEM electrode mounted on the pAT-TPC.

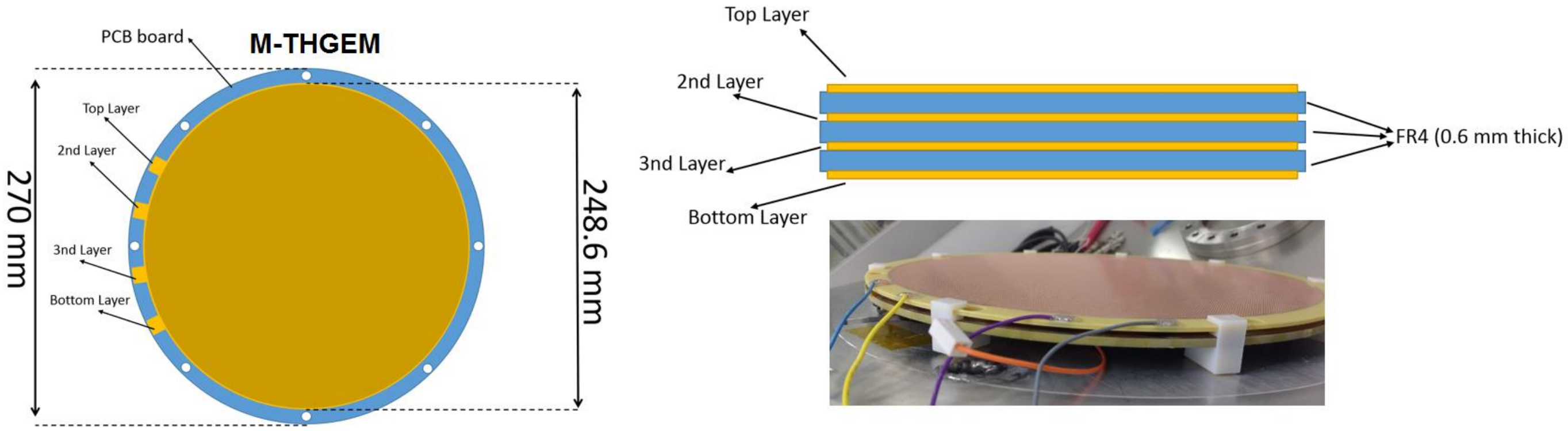

Figure 7. A schematic drawing (left and top right) and a photograph (bottom right) of the M-THGEM mounted on top of the pAT-TPC readout.

The detector was irradiated with 5.5 MeV alpha particles from a collimated 241-Am source, placed behind the thin entrance window. The alpha particles travel along the symmetry axis of the field cage and induce track signals from ionized gas molecules, with time lengths of typically a few tens of μs, depending on the gas pressure (which affects the range of the alpha particles) and on the drift field strength (which modifies the electron drift velocity). At pressures below 300 torr, the alpha particles cross the entire pAT-TPC active volume depositing only a fraction of their total kinetic energy in the gas. At higher pressures (above 300 torr), the alpha particles stop in the pAT-TPC volume, releasing their full energy. The M-THGEM signals were processed with a charge-sensitive preamplifier (ORTEC Model 109A) and the Faster module [24]. The latter includes signal digitization, pulse-shaping (set to typically 33 μs time constant), peak sensing and hold function processing, and data recording.

Figure 8 illustrates an example of the energy spectrum measured in He/(10%)$CO_2$ (black graph) and its comparison to the simulation computed by a Monte Carlo calculation using MCNPX (blue graph) [25]. The sharp peak corresponds to the energy released by particles emitted in the forward direction; they cross the entire field cage all the way to the end-cap readout plane – the total kinetic energy released in the gas under this condition is around 2.4 MeV, with a measured energy dispersion of 5% (compared to the 4.5% computed with MCNPX simulations). The low-energy tail of the spectrum corresponds to energies released by alpha particles emitted at larger angles with respect to the cylindrical axis; they exit the detector active volume and stop on the external walls of the field cage. The difference between the computed and measured intensities of the low-energy tail are due to the uncertainties on reproducing the exact geometry of the collimator, placed in front the 241-Am source.

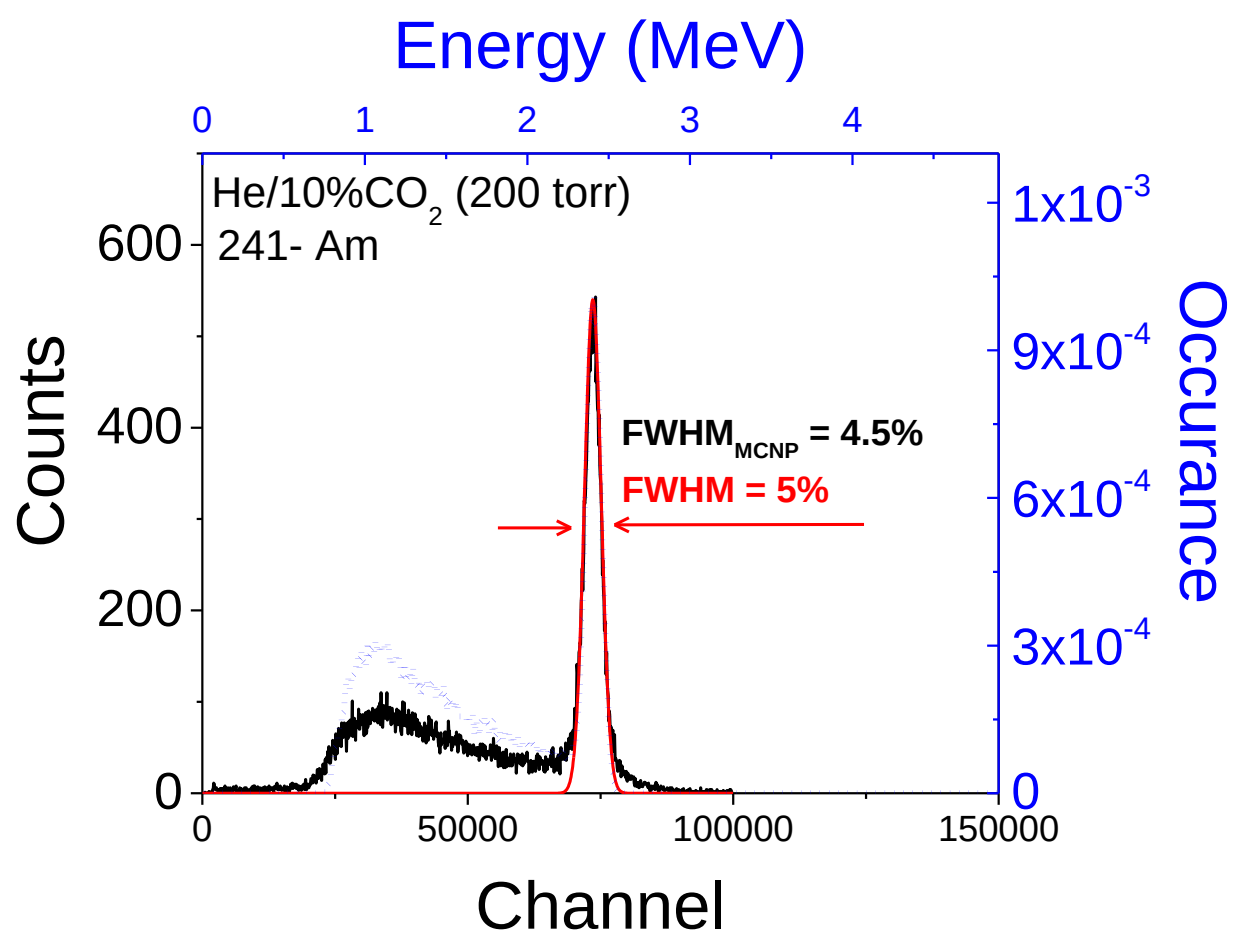

Figure 8. Spectrum of a 5.5 MeV alpha particle crossing the pAT-TPC active volume, recorded with the three-layer M-THGEM readout operated in He-based mixture (with 10% $CO_2$). The experiment (black curve) is compared to a MCNPX simulation (blue curve).

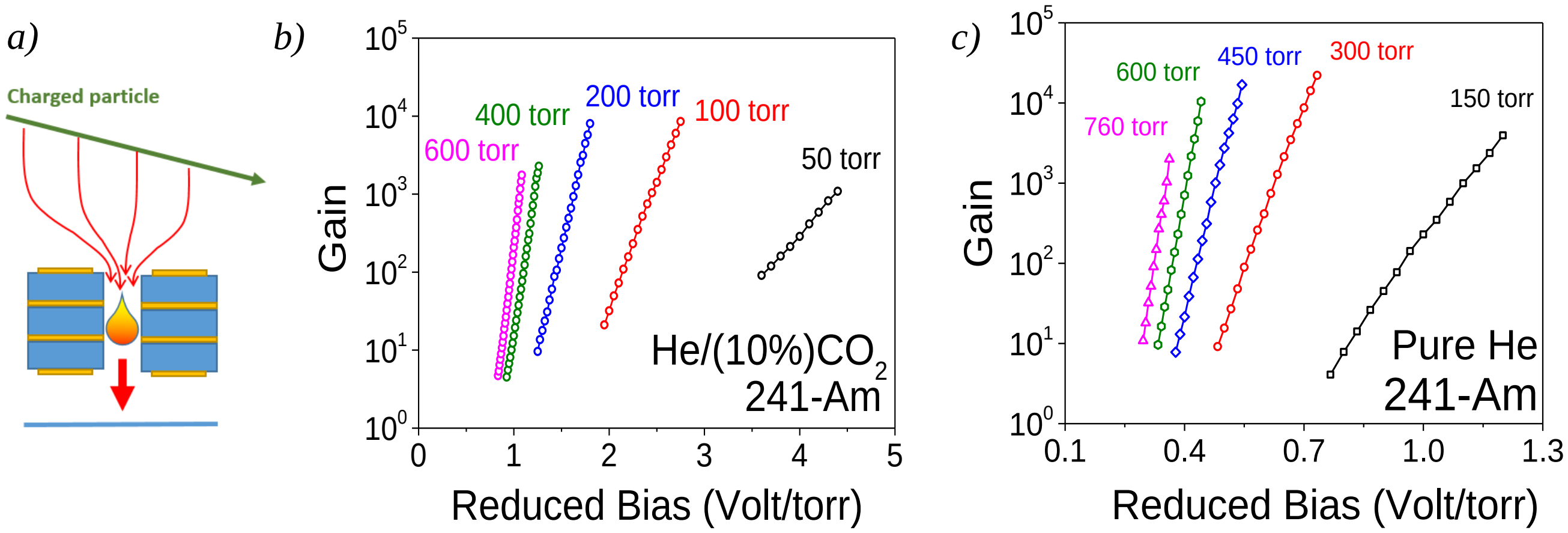

Figure 9. Gain measured with the pAT-TPC readout by a three-layer M-THGEM (schematically drawn in part a), irradiated by a 241-Am source (5.5 MeV alpha), in He/(10%)$CO_2$ (part b) and in pure He (part c).

Similarly to the study performed for the two-layer structure, the characterization of the three-layer M-THGEM detector (Figure 9a) has been carried out in He/(10%)$CO_2$ (Figure 9b) and in pure He (Figure 9c), at different pressures. High-gain operation and relative good energy resolution (estimated of around 2% for 5.5 MeV alpha particles) have been obtained in both gases. The decrease of maximum achievable gain at high pressure is due to the smaller range of the alpha particles, which results in denser ionizations per unit of time, so that the Raether limit condition, i.e. the limit to the multiplication in an electron avalanche, is met at a lower gain. Conversely, the detector shows quite large instabilities at very low pressure (below 100 torr) when relative high voltage is applied to the M-THGEM electrodes, leading to a maximum achievable gain was few times smaller.

As already discussed in the introduction (section 2), asymmetric bias operation is an interesting option bringing additional advantages: the first multiplier stage acts as collector of positive charges, limiting the ion backflow to the drift region. In addition, the avalanche is squeezed towards the lower region of the M-THGEM holes, preventing the occurrence of photon-mediated feedback effects originating from the top electrode surface. Figure 10a illustrates a typical electric field map of the three-layer M-THGEM symmetrically biased (ΔV = 300 V) computed using the ANSII Maxwell software, and compared to the asymmetric bias configuration in which the first potential difference was reduced by a factor of 3 (70% in figure 10b). The gain curves of the M-THGEM asymmetric configuration, measured in 200 torr He/10%$CO_2$ (Figure 10c), shift towards higher reduced biased but reach a similar maximum achievable gain (above $10^4$).

While full electron collection efficiency (Figure 11a) is reached already at a low hole-to-drift field ratio of 20, which corresponds to a gain of about a few hundred, the asymmetric configuration shows a slightly better collection of the ions on the M-THGEM top electrodes, reducing the ion backflow. Both graphs were computed using Garfield simulations. An even higher efficiency on reducing ion backflow, below 10%, is expected in detectors with multi-cascade elements.

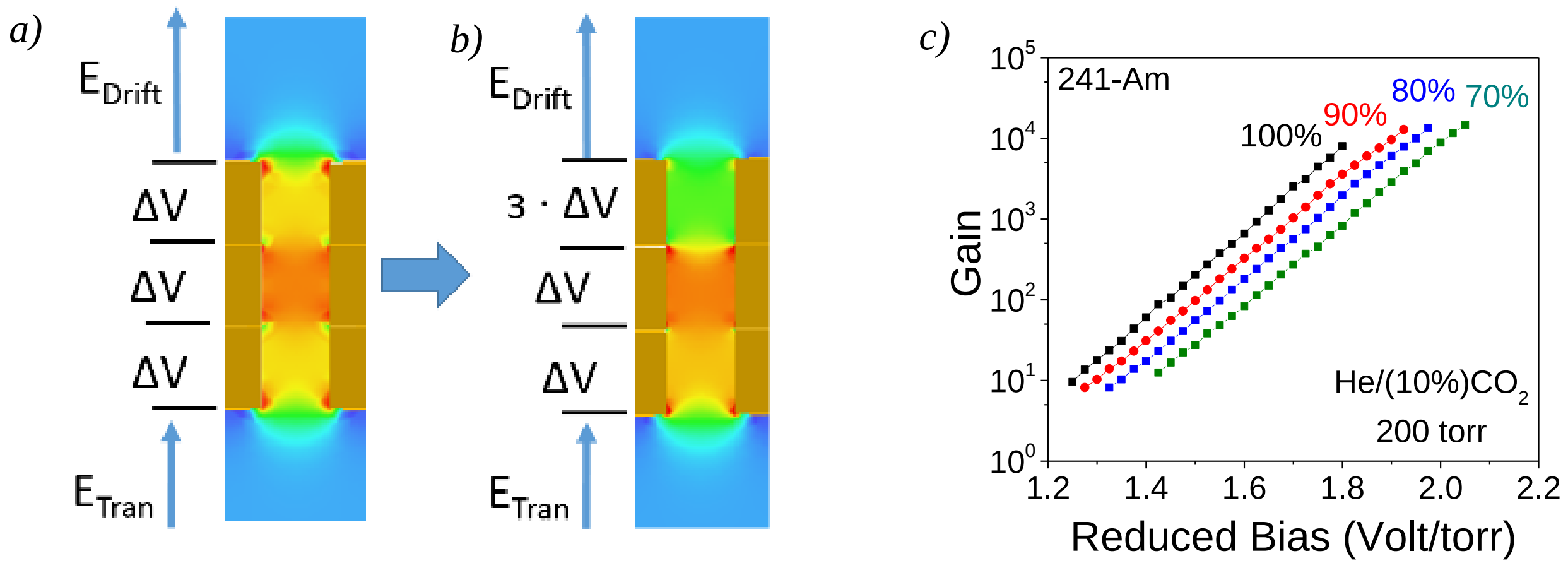

Figure 10. Comparison of the electric field map for a three-layer M-THGEM hole symmetrically biased (part a) and asymmetric biased (part b). 3 represents the reduced fraction of the applied potential to the first multiplier stage. Part c) illustrates measurements of effective gain in 200 torr He/10%$CO_2$ for an asymmetric biased three-layer M-THGEM (black graph), compared to the asymmetric bias configuration for 3 equal to 90% (red graph), 80% (blue graph) and 70% respectively.

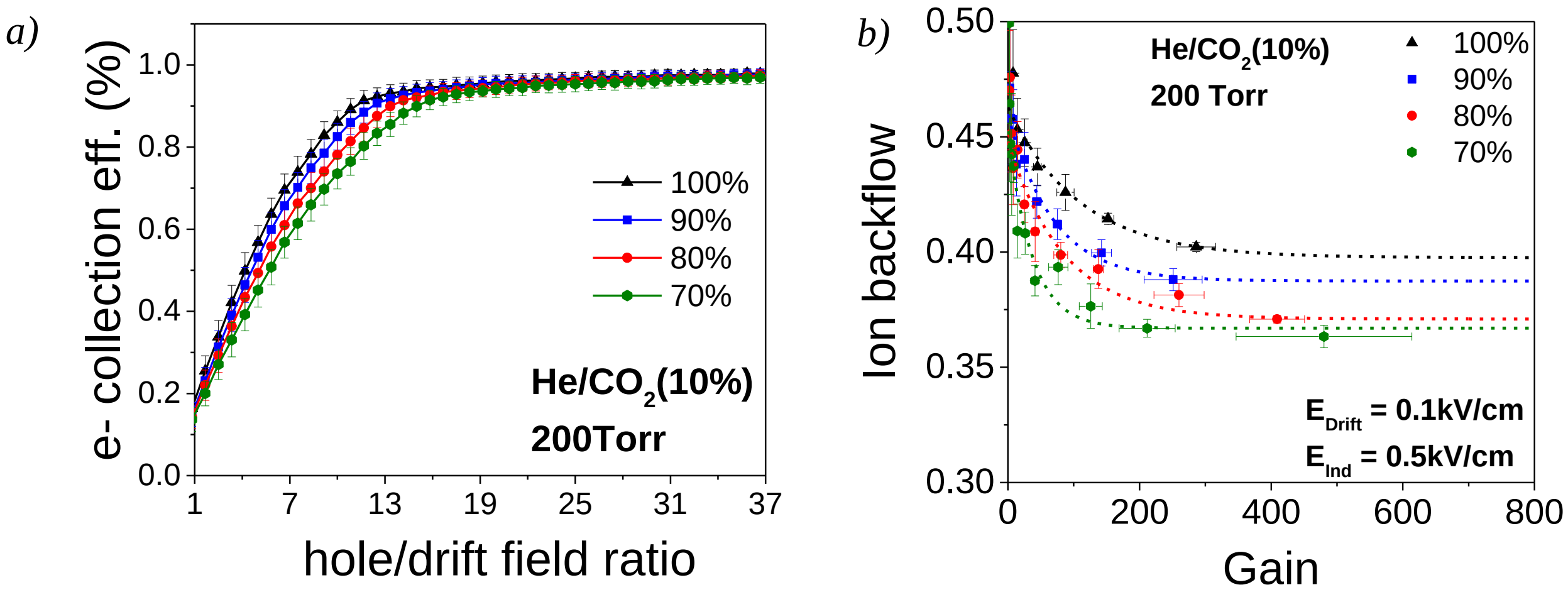

Figure 11. Comparison of the electron collection efficiency (part a) and ion backflow (part b), computed by Garfield simulation for symmetric- (black graph) and asymmetric-mode of operation (i.e. 90% blue graph, 80% red graph and 70% green graph). The percentages indicate the reduced fraction of potential bias applied to the first stage.

## IV. CONCLUSION

The structure and operation mechanism of a new hole-type gas avalanche multiplier, the multi-layer Thick Gaseous Electron Multiplier (M-THGEM), was presented and discussed. The M-THGEM is produced from multi-layer printed circuit boards and consists of a densely perforated assembly of multiple insulating substrate sheets (e.g. FR-4, Kapton, Ceramic, etc.), sandwiched between thin metallic-electrode layers. It is an extremely robust and compact structure, which can cover large areas without supporting frames. The performance capabilities of two detector geometries, the two-layer and three-layer M-THGEM, have been investigated in this work for the first time. A stable operational gain above $10^5$ and $10^6$ has been obtained for the two-layer M-THGEM, respectively with single- and double-element configuration, in a He-based mixture (10% $CO_2$). The same gain has been measured in pure He under the same experimental conditions. In similar gases, a large area three-layer M-THGEM was characterized using 5.5 MeV alpha particle tracks, reaching a maximum achievable gain above $10^3$-$10^4$ depending of the gas pressure. Further extensive experimental studies are in preparation and will be reported elsewhere; these will include the investigation of the detector response uniformity over a large area, energy resolution, ion backflow suppression capability in symmetric and asymmetric modes of operation, long-term gain stability, effective gain in different gas mixtures, spatial resolution and counting rate capability.

The M-THGEM has several potential advantages compared to the traditional multi-cascade hole-type (THGEM/GEM) detector, including negligible loss of charge from one multiplication stage to the other, which leads to a higher effective gain at lower operational voltage. The avalanche confinement within the holes leads to a significant reduction of photon-mediated secondary effects, resulting in a more stable high-gain operation in pure noble gas, without needs of a quenching gas. This is particularly efficient in the case of the three-layer M-THGEM structure, due to the peculiar geometrical configuration of the electric dipole field, inherently compressed within the most inner multiplication stage. One possible drawback of the M-THGEM detector is a somewhat higher ion-backflow, which can be mitigated by operating the structure with an asymmetric bias configuration. When operated at sufficiently high gain, the asymmetric mode of operation reaches full electron collection efficiency, and no loss of energy resolution is expected. The bias potential provided by the inner electrodes has the advantage of stabilizing the electric field geometry in the avalanche region, leading to a substantial mitigation of the insulator substrate charging up and a stable gas gain over time. Finally, because of the compact assembly of the many multiplication stages, the avalanche region is distributed over a lengthy volume, several times longer than the electron mean free path, and a high gain can also be achieved at an exceptionally low pressure. To increase the localization capability, the M-THGEM can be combined with other high-granularity MPGD structures (i.e. Micromegas, μPIC, etc.) in a hybrid readout configuration, performing as a pre-amplification stage.

The M-THGEM was specifically conceived for applications requiring high-gain at low pressure such as heavy-ion detection for tracking/triggering, as well as operation in pure noble gas, such as needed for time-projection chamber readout in active target mode. Other potential applications may include large-area UV photon detectors, muon trackers, hadron calorimetry, X-ray/neutron imaging and secondary scintillation readout for rare event physics.